\documentclass[aps,twocolumn,superscriptaddress,showpacs,floats]{revtex4}
\usepackage{amsmath}
\usepackage{amsfonts}
\usepackage{amssymb}
\usepackage{euscript}
\usepackage{bm}
\usepackage[dvips]{graphicx}

\begin{document}

\title{Geometric Symmetries in Superfluid Vortex Dynamics}

\author{Evgeny Kozik}
 \affiliation{Institute for Theoretical Physics, ETH Zurich, CH-8093 Zurich, Switzerland}
\author{Boris Svistunov}
\affiliation{Department of Physics, University of Massachusetts,
Amherst, MA 01003} \affiliation{Russian Research Center
``Kurchatov Institute'', 123182 Moscow, Russia}

\begin{abstract}
Dynamics of quantized vortex lines in a superfluid  feature  symmetries associated with the geometric character of the  complex-valued field, $w(z)=x(z)+iy(z)$, describing the instant shape of the line. Along with a natural set of Noether's constants of motion, which---apart from their rather specific expressions in terms of $w(z)$---are nothing but  components of the total linear and angular momenta of the fluid, the geometric symmetry brings about crucial consequences for kinetics of distortion waves on the vortex lines---the Kelvin waves. It is the geometric symmetry that renders Kelvin-wave cascade local in the wavenumber space. Similar considerations apply to other systems with purely geometric degrees of freedom.
\end{abstract}


%
%
%
%
%

\maketitle

Throughout various fields of physics, symmetries form the very basis of the analytical approach to problems. Besides immediately providing one with the corresponding integrals of motion, in many cases symmetries allow to avoid lengthy calculations, a famous example being the selection rules in atomic physics. For a wide class of systems described in terms of purely geometrical fields of displacements---such as liquid crystals, waves on strings, surfaces and membranes---geometrical symmetries affect dynamics in a special way. Quantized vortices in superfluids are a typical example of such systems.

Superfluid turbulence (ST) is a random or polarized tangle of quantized vortex lines  \cite{Donnelly, Vinen06}. Fundamentally interesting and especially non-trivial in terms of relaxation dynamics is the case of zero temperature intensively studied during last one and a half decade \cite{Sv95, Vinen2000,  Davis, Tsubota00, Vinen2001, Kivotides, Vinen_2003, KS_04, KS_05, KS_05_vortex_phonon, Lvov, KS_crossover, Lvov2, KS_scan, KS_JLTP}, with impressive recent experimental achievements \cite{Bradley, Golov,Golov2,Helsinki,Weiler,Henn}. It has been recognized that Kelvin waves (kelvons)---distortion waves on quantized vortex filaments in a (super)fluid---play a crucial role in the ST decay at $T=0$ by providing the only mechanism of energy transfer \cite{Sv95,Vinen_2003, KS_04} to shorter length scales, where the dissipation due to phonon emission becomes efficient \cite{Vinen2000, KS_05_vortex_phonon}. Recently, physics of the Kelvin-wave cascade has become one of the central topics in ST research \cite{KS_05, Lvov, KS_crossover, Lvov2, KS_scan, LN_nonlocal, LN}.

In the theory of the Kelvin-wave cascade (see Ref.~\cite{KS_JLTP}, and references therein), geometrical symmetries impose fundamental constraints on kelvon kinetics. They are usually discussed in connection with the kinematics of kelvon collisions dictating the absence of inelastic kelvon scattering processes and the fact that the leading kinetic channel is the three-kelvon scattering. As important as kinematics, however, is the question of
\textit{locality} (or its absence) of the kelvon collisions in the wavenumber space. The locality implies that the main contribution to energy exchange at a given wavenumber scale is mostly due to scattering of kelvons with wavenumbers of the same order. If the locality is established, the cascade spectrum immediately follows from kinematics and a basic analysis of dimensions \cite{KS_04}. However, recent Ref.~\cite{LN_nonlocal}  claimed kelvon kinetics to be essentially \textit{non-local}, declaring inadequacy of a wealth of theoretical results based on the locality;  an alternative non-local theory of the Kelvin-wave cascade was then put forward in Ref.~\cite{LN}.

In this Letter, we explore additional symmetries of vortex dynamics that follow from the geometric nature of the field describing the position of the vortex line.  We show that these symmetries impose further constraints on kinetics;  in particular, with respect to coupling between waves of drastically different length scales. It is crucially due to these symmetries that the generic non-local scenario of Refs.~\cite{LN_nonlocal, LN} fails, yielding to the specific (and thus non-trivial) local one.  Our approach applies to geometric fields of an arbitrary nature.

We start with a mathematical formulation of the problem of Kelvin-wave dynamics on a single vortex line in the limit of zero temperature---when the normal component is negligible. In this case, a convenient dynamical variable is the complex geometrical field $w(z,t)= x(z,t)+iy(z,t)$, parameterized by the time $t$ and the $z$-coordinate of the Cartesian coordinate system such that $x(z_0,t)$ and $y(z_0,t)$ are the coordinates of the vortex-line crossing with the plane $z=z_0$. In terms of $w(z,t)$, the equations of motion take on a Hamiltonian form \cite{Sv95},
\begin{equation}
i\dot{w} = {\delta H\over \delta w^*} ,
\label{eq_motion}
\end{equation}
\begin{equation}
H= (\kappa/4\pi)  \int  {\left[ 1+{\rm Re} \,
w'^*(z_1)w'(z_2) \right]  dz_1 dz_2 \over \sqrt{ (z_1-z_2)^2+|w(z_1)-w(z_2)|^2}} \, ,
\label{ham}
\end{equation}
where $\kappa$ is the fluid-specific quantum of circulation and the integration is along the whole vortex line.
The integral in the r.h.s. of Eq.~(\ref{ham}) is singular at $(z_1-z_2) \to 0$ and requires a regularization procedure at distances of order of the vortex core radius $a_0$ described in detail in Ref.~\cite{KS_JLTP}. Clearly, Eq.~(\ref{ham}) is well-defined only in the case when the function $w(z)$ is single-valued, which is the case for the problem of Kelvin-wave cascade where the amplitudes of distortions get progressively smaller down the length scales \cite{KS_04}, provided the $z$-axis is chosen along the direction of the unperturbed vortex line. The condition of smallness of Kelvin-wave amplitudes relative to the wavelengths is given by the small parameter $\alpha(z_1,z_2) = |w(z_1)-w(z_2)| / |z_1-z_2| \ll 1$ allowing one to use the language of normal modes, kelvons. The kelvon dispersion follows from the standard (bi-)linearization of the Hamiltonian (\ref{ham}) with respect to $\alpha \ll 1$ and the Fourier transform $w(z)=L^{-1/2}\sum w_k \exp(ikz)$, $L$ being the linear system size in $z$-direction. Correspondingly, the higher-order non-linear terms are regarded as kelvon scattering (for a detailed review, see \cite{KS_JLTP}).

Let us now turn to implications of the six geometrical symmetries. The translational symmetry along the $z$-axis is responsible for the conservation of the total momentum of kelvons (equal to the kelvon contribution to the $z$-component of the total momentum of the liquid):
\begin{equation}
P \propto \int dz \, w^* w' \, .
\label{P}
\end{equation}
The symmetry with respect to the rotations around the $z$ axis---which in our representation is nothing but the global U(1) symmetry---implies the conservation of the total number of kelvons,
\begin{equation}
N \propto \int dz \, |w|^2\, ,
\label{N}
\end{equation}
which is proportional (with the negative sign) to the kelvon contribution to the $z$-component of the angular momentum.
Despite their geometric origin, the above two symmetries are quite generic: The translational symmetry is  standard in the theory of fields of arbitrary nature, and the U(1) symmetry of complex-valued fields is very common. More specific symmetries, based essentially on the geometric nature of $w(z)$, are the translations in the direction
of a unit vector $\hat{n} = (n_x,n_y,0)$ perpendicular to the $z$-axis, and the rotations around $\hat{n}$.  In what follows we will refer to those symmetries as the {\it shift} and {\it tilt} symmetries, respectively.

The  shift  of the line by the distance $l$ is
\begin{equation}
w(z) \, \to\,  w(z) + n l  \, ,
\label{shift}
\end{equation}
where $n=n_x+in_y$. Invariance of the Hamiltonian with respect to (\ref{shift}) implies $n^* \int dz \, \dot{w} + {\rm c.c. } = 0 $, and via linear independence of $n$ and $n^*$ means the conservation of
\begin{equation}
P_\perp \, =\, \int dz \,  w  \, .
\label{P_perp}
\end{equation}
Up to a dimensional factor, $P_\perp$ is the  $xy$-component of the momentum of the fluid. In the kelvon terminology,
the quantity $P_\perp$ is, up to a normalization, the amplitude of kelvon condensate. The  conservation of $P_\perp$ means that the kelvon
condensate does not interact with the rest of the system, and vice versa: kelvon dynamics are not sensitive
to the condensate amplitude. This fact---by no means surprising since the condensate is a mere shift
of the whole vortex line---proves crucially important for the structure of the kelvon kinetic processes (see below).

An infinitesimal tilt by the angle $\delta \phi$ is expressed by $w(z) \to   w(z) -w'(z) \delta z + \delta w $
with  $\delta w \equiv \delta x + i \delta y$ and $\delta z$ obeying $ \delta {\bf r}\, =\, [\hat{n}\times {\bf r}] \, \delta \phi $,
where $ \delta {\bf r} = (\delta x, \delta y, \delta z)$.
Observing that $\delta z  = (i/2) (w^*n-wn^*) \,  \delta \phi $, $\delta w =-izn \delta \phi$,
we write the tilt transformation as
\begin{equation}
w(z) \, \to\,  w(z)  - i \left[  w' (w^*n-wn^*)/2 + zn \right]  \delta \phi   \, .
\label{tilt_final}
\end{equation}
Invariance of the Hamiltonian with respect to (\ref{tilt_final}) implies $\int dz \, \dot{w}^* \left[ w' (w^*n-wn^*) + 2zn \right] + {\rm c.c. }=0$
and, by linear independence of $n$ and $n^*$, leads to the constant of motion
\begin{equation}
L_\perp \, =\, \int dz \,  \left[2 w z  - |w|^2 w' \right]  \, ,
\label{L_perp}
\end{equation}
which, up to a dimensional factor, is the kelvon contribution to the the  $xy$-component of the angular momentum.

Due to the conservation of $P$, $N$, and energy, and the fact that our system is one-dimensional, the leading kinetic channel is the three-kelvon {\it elastic} scattering (i.e., the 6-wave process, in the non-linear physics terminology). This allows one to write down a general form of the kinetic equation \cite{KS_04,KS_JLTP}, $\dot{n}_1={\rm Coll}_{k_1}$, 
\begin{equation}
{\rm Coll}_{k_1}[n_k]\! =  \!  \sum_{k_2, \ldots ,
k_6} \! \! \!  \! |V_{1,2,3}^{4,5,6}|^2  \, \delta(\Delta
\omega) \, \delta(\Delta k) \left( f_{4,5,6}^{1,2,3} \! -
\! f_{1,2,3}^{4,5,6} \right )  , \label{KE}
\end{equation}
where $n_i \equiv n_{k_i}$ are the occupation numbers of kelvons with momenta $k_i$ [related to the Fourier transforms of $w(z)$ via $n_k= (\kappa \rho/2\hbar) \langle w^*_k w_k\rangle$, $\rho$ being the fluid density], $\Delta k =
k_1+k_2+k_3-k_4-k_5-k_6$, $\Delta\omega =  \omega_1+\omega_2+\omega_3-\omega_4-\omega_5-\omega_6$, $\omega_k$ is the kelvon dispersion, and $f_{1,2,3}^{4,5,6}=n_1n_2n_3(n_4n_5+n_4n_6+n_5n_6)$. Only the effective three-kelvon vertex $V_{1,2,3}^{4,5,6}$ needs to be derived from Eq.~(\ref{ham}).

If the momentum-space locality of kelvon collisions is proven, i.e. if the main contribution to ${\rm Coll}_{k_1}$ comes from wavenumbers of order $k_1$, then the cascade spectrum,
\begin{equation}
n_k \, \propto\, k^{-17/5}  \, ,
\label{17_5}
\end{equation}
follows by a straightforward dimensional analysis of Eq. (\ref{KE}) with $V_{1,2,3}^{4,5,6} \propto k_1^6$ \cite{KS_04,KS_JLTP}. In our Ref.~\cite{KS_04}, the proof of locality was done numerically. Subsequent direct simulation \cite{KS_05} of the dynamic model (\ref{eq_motion})-(\ref{ham}) revealed perfect agreement with the local spectrum (\ref{17_5}). However, Ref.~\cite{LN_nonlocal}  claimed non-locality and thus irrelevance of the spectrum (\ref{17_5}), the agreement with numeric simulation being attributed to the lack of sufficient resolution. Mathematically, non-locality of a given solution manifests itself as a divergence of the collision integral (\ref{KE}). For the power-law solution (\ref{17_5}), the scaling suggests divergence at the lower limit---when one or more momenta are much smaller than $k_1$---which has to be removed by exact cancellations if the solution is local.  The  conclusion of Ref.~\cite{LN_nonlocal} is based on a direct evaluation of the integrals, which suggests the absence of such cancellations, and thus should be taken seriously. The situation is complicated by the following circumstances: (i) the integrals are rather cumbersome \cite{KS_04,KS_JLTP};  (ii) a technical mistake made in Ref.~\cite{KS_04} (but fixed in Ref.~\cite{KS_JLTP}) renders the numeric analysis of Ref.~\cite{KS_04} not trustworthy; (iii) the form of the integrals proposed in Ref.~\cite{LN_nonlocal} is {\it different} from what we find in Ref.~\cite{KS_JLTP}, so that a direct evaluation of the integrals does not yet resolve the controversy. Under these circumstances, it becomes extremely desirable to address the issue of locality without resorting to details of the integrals in the collision term, and the geometric symmetry yields this opportunity.

We observe that even when the leading contributions to the collision term are due to interactions between dramatically different wavenumber scales, the kinetics is  still {\it semi}-local in the sense that interactions between short waves are coupled to an effective {\it quasi-static external potential} formed by the slow long-wave modes. Physically, the semi-locality is dictated by the separation of time scales of the short- and long-wave dynamics with the observation that the typical dynamical correlation length at a given short scale $\lambda$ is of order $\lambda$. Thus, the short-wave fast processes can, in principle, only couple to \textit{spatially local} instantaneous characteristics of the slow modes, such  as their amplitudes and spacial derivatives.  Our central point is that, for fields of purely geometrical nature considered here, the tilt and shift symmetries rule out coupling to the amplitude of the long-wave field and its first derivative, the angle. This enforces the key constraint on the asymptotic form of the integrand in the collision term (\ref{KE}) when one of $k_i$, say $k_2$, is much smaller than $k_1$: the wavenumber $k_2 \ll k_1$ can only enter the collision term through the \textit{curvature} of the long-wave components or higher derivatives. On the other hand, for the spectrum (\ref{17_5}), the curvature is actually due to \textit{large} wavenumbers $\sim k_1$, contributions from $k_2 \ll k_1$ being negligible. This immediately implies that ${\rm Coll}_{k_1}$ builds up exclusively at $k_2, \ldots, k_6 \sim k_1$ and is therefore local.

To formalize this proof, with providing  an explicit insight, we consider an equation of motion for short-wave distortions on a curved vortex line. Our starting point is the general Biot-Savart equation (BSE) for the radius-vector ${\bf s}$ of an element of a single vortex line, conveniently associated with the Lagrangian functional
\begin{equation}
L= \frac{2}{3}\! \int\! \!  \left( \bigl[{\bf \dot{s}} \! - \! {\bf v}({\bf s}) \bigr]\!  \times \! {\bf s} \right) \! \cdot \! d {\bf s} - \! E , \; \; E= \frac{\kappa}{4 \pi}\!  \int \! \frac{d {\bf s}\!  \cdot\!  d {\bf s}_0 }{ |{\bf s}\!  - \! {\bf s_0}| }, \label{Lag}
\end{equation}
where the effect of the rest of the tangle is taken into account by the external velocity field ${\bf v({\bf r})}$, and $E$ is nothing but Eq.~(\ref{ham}) in terms of ${\bf s}$. We split ${\bf s}$ into the long-wave part ${\bf g}$ with the typical curvature radius $R_0$ and the vector ${\bf w}$ of small  short-wave distortions of wavelengths $\sim \lambda \ll R_0$ , ${\bf s}={\bf g}+{\bf w}$. Due to the Galilean invariance, the short-wave dynamics is insensitive to ${\bf v}$, which varies at the scales of order (and larger than) the inter-vortex distance $l_0 \gg \lambda$, whereas for ${\bf g}$ one can distinguish two regimes \cite{KS_crossover}: (a) ${\bf \dot{g}} \approx {\bf v}({\bf g})$ if $R_0 \gtrsim l_0$ and there is a large-scale polarization of the tangle, and (b) ${\bf \dot{g}} \approx \beta \, {\bf g}' \times {\bf g}'' $ with $ \beta= (\kappa/4\pi) \ln(R_0/a_0)$---the local induction approximation (LIA) \cite{Donnelly}---if there is no large-scale flow or $l_0 \gg R_0 \gg \lambda$ meaning that the velocity field ${\bf v}$ is negligible. For the question of locality of the solution (\ref{17_5}), $R_0$ needs to be the largest scale of the pure Kelvin-wave cascade---the scale at which LIA dominates the dynamics and Kelvin waves are generated by reconnections---the case (b). To the leading approximation, dynamics of ${\bf g}$ are insensitive to the short-wave structure ${\bf w}$ (c.f. \cite{Sv95}) since the long-wave field plays the role of a condensate amplitude with respect to short-wave dynamics conserving the condensate in view of Eq.~(\ref{P_perp}). Taking into account that the LIA conserves the line length, it is consistent to parameterize the fields ${\bf g}$ and ${\bf w}$ by the arc length $\xi$ of the filed ${\bf g}$. In correspondence with the field $w(z)$, we require that ${\bf w}(\xi)$  lie in the plane perpendicular to the local tangent ${\bf g}'(\xi)$, ${\bf w}(\xi) \cdot {\bf g}'(\xi)=0$.  Introducing a local basis attached to the curve ${\bf g}(\xi)$, ${\bf e}_3={\bf g}'$, ${\bf e}_1={\bf g}''/|{\bf g}''|$, ${\bf e}_2={\bf e}_3\times{\bf e}_1$, so that ${\bf w}(\xi)=x(\xi) {\bf e}_1(\xi) + y(\xi) {\bf e}_2(\xi)$, we define $w(\xi)\equiv x(\xi)+iy(\xi)$, which for Eq.~(\ref{Lag}) yields 
\begin{equation}
L=\int d\xi \, [1- g''(w+w^*)/3\,] \,(i\dot{w}w^* - i\dot{w}^*w )/2\; - \; E, \label{Lag_g}
\end{equation}
%
%
\begin{widetext}
\begin{equation}
E= \frac{\kappa}{4 \pi} \int d\xi_1 \, d\xi_2  \frac{1 + {\rm Re} \left[ (\xi_1-\xi_2) (g''_2 w'_2 - g''_1 w'_1) - g''_2 w_2 - g''_1 w_1 + w'^{*}_1 w'_2 \right] }{\sqrt{ \left[ 1+ g''^{2}_1 (\xi_1-\xi_2)^2/4 - {\rm Re}(g''_1 w_1 + g''_2 w_2) \right] (\xi_1-\xi_2)^2 + |w_1-w_2|^2 }} \,, \label{ham_g}
\end{equation}
\end{widetext}
where $g''_{1,2} \equiv |{\bf g''} (\xi_{1,2})|$ and $w_{1,2}\equiv w(\xi_{1,2})$. In deriving (\ref{ham_g}), we replace ${\bf g}$ by its Taylor series expansion to the first nonvanishing order, justified by the small parameter $\lambda/R_0 \ll 1$. Note that ${\bf g}$ and ${\bf g}'$ explicitly drop out due to the shift and tilt symmetries leaving the first non-vanishing contribution to be proportional to the curvature $g''$. In the limit $g'' \to 0$, we recover Eqs.~(\ref{eq_motion}), (\ref{ham}). Equations (\ref{Lag_g}) and  (\ref{ham_g}) substantiate our central observation prescribed by the symmetry---the leading contribution to Kelvon dynamics from coupling to long-wave modes is proportional to the large-scale curvature $g''$. 

By the standard procedure, expanding with respect to $\alpha \ll 1$ and $g''|w_{1,2}|  \ll 1$, and introducing $w(\xi)=\mathcal{L}^{-1/2}\sum w_k \exp(ik \xi)$ with $\mathcal{L}$ the total line length, we observe that the leading kinetic processes coupled to $g''$ are five-wave in terms of the fast components $w_k$, which results in the non-local contribution to the collision term
\begin{equation}
{\rm Coll}^{\mathrm{(nl)}}_{ k_1}[n_k] =  \!  \overline{(g'')^2} \sum_{k_2, \ldots ,
k_5} \! \! \!  \! |\tilde{V}(\{k_i\})|^2  \, \delta(\Delta
\omega) \, \delta(\Delta k)  \tilde{f}(\{n_i\}), \label{coll_nl}
\end{equation}
where $\overline{(g'')^2}=\mathcal{L}^{-1}\int \langle (g'')^2(\xi) \rangle d\xi$; $\tilde{V}$ and $\tilde{f}$ are some functions such that $\tilde{V}(\sigma k_1, \ldots, \sigma k_5)=\sigma^4 \tilde{V}(k_1, \ldots, k_5)$, and $\tilde{f}(\sigma n_1, \ldots, \sigma n_5)=\sigma^4 \tilde{f}(n_1, \ldots, n_5)$. For the fast field with the typical wavenumbers $ k_i \sim k$, Eq.~(\ref{coll_nl}) yields ${\rm Coll}^{\mathrm{(nl)}}_{ k}\sim\overline{(g'')^2} \; k^9 n^4_k$, whereas the local contribution follows from Eq.~(\ref{KE}): ${\rm Coll}^{\mathrm{(l)}}_{ k}\sim k^{14} n^5_k$. The derivation of Eq.~(\ref{coll_nl}) requires two conditions to be satisfied: (i) typical kinetic rates for the modes $w_k$ should be much smaller than the corresponding frequencies $\omega_k \propto k^2$, which allows to use the Wick's theorem upon averaging the dynamic equations, and (ii) the Knudsen limit for the fast field, i.e. the mean free path of the fast modes much longer than the wavelengths of the slow field ${\bf g}$, enabling averaging over $\xi$. Both conditions are met for the spectrum (\ref{17_5}) in question and large $k$. 

Equation (\ref{coll_nl}) reveals the major inconsistency of the assumption of considerable coupling between different length scales: for the spectrum (\ref{17_5}), the mean-square curvature builds up at \textit{large} wavenumbers, $\overline{(g'')^2} \propto k^{8/5}$, meaning that the leading ``non-local'' term ${\rm Coll}^{\mathrm{(nl)}}_k$ is actually \textit{local}. In real tangles, where dynamics at large scales can be rather complex \cite{KS_crossover}, it is relevant to estimate the correction to short-wave kinetics due to the large-scale structure. The answer follows from Eq.~(\ref{coll_nl}) if we restrict the spectrum of ${\bf g}$ to $k\sim R_0^{-1}$:  ${\rm Coll}^{\mathrm{(nl)}}_k/{\rm Coll}^{\mathrm{(l)}}_k \propto R_0^{-2}/(k^5 n_k)$, and thus, at $k \gg R_0^{-1}$, where the pure Kelvin-wave cascade is expected to develop, any coupling to the large scales quickly vanishes.

It is important to emphasize that the conclusion about non-locality of the spectrum (\ref{17_5}) made in Ref.~\cite{LN_nonlocal} unquestionably applies to the differential model---the local (in real space) nonlinear equation (LNE)---introduced and studied there as a ``simplification'' of Eqs.~(\ref{eq_motion}), (\ref{ham}). That is because, as opposed to (\ref{ham}), LNE {\it does not} respect the tilt symmetry. This fact, however, means nothing but irrelevance of LNE to the problem of the Kelvin-wave cascade. In their evaluation of the effective vertex, the authors of Ref.~\cite{LN_nonlocal} claimed revealing some relevant terms  missed in our Ref.~\cite{KS_JLTP}. In view of the explicit tilt symmetry violation by the results of Refs.~\cite{LN_nonlocal, LN}, introducing these terms is likely to be the main source the mistake. More generally, uncontrolled approximations and errors \textit{generically} lead to the non-local L'vov-Nazarenko cascade with the $n_k \propto k^{-11/3}$ \cite{LN} spectrum.

As another example of a system characterized by the crucial effect of geometric symmetries on non-linear dynamics, we mention smectic liquid crystals \cite{smectics} where the tilt symmetry is known to be responsible for non-trivial non-linerar effects. The relationship between geometric symmetry and locality of weak-turbulent spectra found in the context of Kelvin waves can thus prove important for other systems with geometric degrees of freedom.

We are grateful to Christian Santangelo for introducing us to the crucial role played by geometric symmetries in the theory of smectic liquid crystals.

{\it Note added.--} After our geometric argument of locality had been presented at the Symposium on Superfluids
under Rotation (Lammi, Finland, April 2010), Lebedev and L'vov [arXiv:1005.4575] attempted to build a counter-argument. We comment
on this work in arXiv:1006.1789. First-principles simulations clearly resolving the local Kelvin-wave spectrum from the L'vov-Nazarenko one were
performed by us recently [arXiv:1007.4927].

\end{document}